\documentclass[12pt,twocolumn]{emulateapj}
\usepackage{apjfonts}

\def\lsim{\lower 2pt \hbox{$\, \buildrel {\scriptstyle <}\over
{\scriptstyle \sim}\,$}}

\slugcomment{To appear in ApJ 2004}

\shorttitle{Formation of Semi-relativistic Jets from Magnetospheres of
  Accreting Neutron Stars}
\shortauthors{Kato, Hayashi, \& Matsumoto}

\begin{document}

\title{Formation of Semi-relativistic Jets from Magnetospheres of
  Accreting Neutron Stars: Injection of Hot Bubbles into a Magnetic
  Tower}

\author{Y. Kato\thanks{Department of Physics, Graduate School of
Science and Technology, Chiba University, Inage-ku, Chiba 263-8522,
Japan}, M. R. Hayashi\thanks{National Astronomical Observatory,
Mitaka, Tokyo 181-8588, Japan}, and R. Matsumoto\thanks{Department of
Physics, Faculty of Science, Chiba University, Inage-ku, Chiba
263-8522, Japan}}

\begin{abstract}
We present the results of 2.5-dimensional resistive
magnetohydrodynamic (MHD) simulations of the magnetic interaction
between a weakly magnetized neutron star and its accretion disk.
General relativistic effects are simulated by using the
pseudo-Newtonian potential.  We find that well-collimated jets
traveling along the rotation axis of the disk are formed by the
following mechanism: (1) The magnetic loops connecting the neutron
star and the disk are twisted due to the differential rotation between
the neutron star and the disk.  (2) Twist injection from the
disk initiates expansion of the loop.  (3) The expanding magnetic
loops create a magnetic tower in which accelerated disk material
travel as collimated bipolar jets.  The propagation speed of the
working surface of the jet is the order of 10\% of the speed of light
($\sim 0.1c$).  (4) Magnetic reconnection taking place inside the
expanding magnetic loops injects hot bubbles intermittently into the
magnetic tower.  The ejection speed of the bubble is the order of the
local Alfv\'{e}n speed of the launching point and $\sim 0.2c$ in our
simulations.  (5) The hot bubbles moving inside the tower catch up
with the working surface of the jet.  High energy electrons created by
the magnetic reconnection are a plausible source of radio emission.
Our model can explain the formation process of a narrow jet from a
weakly magnetized ($|\mbox{\boldmath$B_{*}$}|\le 10^{9}$ gauss)
neutron star and the correlation between radio flares of the core and
of the lobe observed in Sco X-1.
\end{abstract}


\section{INTRODUCTION}
The mechanism of jet formation is one of the most important subjects
of research in astrophysics.  Blandford \& Payne (1982) studied
magneto-centrifugal acceleration along a magnetic field line threading
an accretion disk.  Uchida \& Shibata (1985) and Shibata \& Uchida
(1986) showed  by time-dependent magnetohydrodynamical (MHD)
simulations that bipolar jets are produced from an accretion disk
threaded by open magnetic field lines.

Observational evidence indicate the presence of semi-relativistic jets
in some galactic X-ray binaries (XRBs) such as SS433 (Hjellming \&
Johnston 1981), Cygnus X-3 (Mart{\'{\i}} {\it et al.} 2000,
Mioduszewski {\it et al.} 2001), Sco X-1 (Fomalont {\it et al.}
2001a,b), and Circinus X-1 (Fender {\it et al.} 1998).  The speed of
jets in such XRBs indicates that the jets are launched close to the
last stable orbit of the accretion disk. Therefore, we should include
general relativistic effects around a compact object.  In addition,
the magnetic field of the compact object may play an essential role in
the dynamics of the jets and the accretion flow.  Recent observations
of XRBs by the Very Large Baseline Interferometer (VLBI) revealed
their energetic activities such as radio flares and radio jets.  VLBI
discovered the motions and variabilities of radio lobes in Sco X-1
(Fomalont {\it et al.} 2001a,b).  According to their observations,
radio lobes are ejected from the core of Sco X-1.  After the radio
flare of the core, the intensity of the advancing radio lobe
increases.  The correlation between the core flare and the lobe flare
in Sco X-1 indicates that explosive events at the core transports
energy to the lobe.

A plausible model of advancing lobes is the working surface of jets
ejected from the core.  However, the existence of large scale open
magnetic fields assumed in the magneto-centrifugal model of jet
formation is not evident in XRBs.  Instead, we should take into
account the magnetic interaction between the neutron star and its
accretion disk.

Lynden-Bell \& Boily (1994) studied the evolution of force-free
magnetic loops anchored to the star and the disk.  They obtained
self-similar solutions for the evolution of the magnetic loops.  They
found that the loops are unstable against the twist injection from the
rotating disk and that the loops expand along a direction of 60
degrees from the rotation axis of the disk.  Lovelace {\it et al.}
(1995) pointed out that the dipole magnetic field of the star deforms
itself into an open magnetic field due to the differential rotation
between the star and the disk.  Subsequently, Lynden-Bell (1996;
hereafter referred to as LB96) showed that a cylindrical magnetic
tower is formed when the disk is surrounded by external plasma with
finite pressure.

Hayashi, Shibata, and Matsumoto (1996; hereafter referred to as HSM96)
carried out the first numerical calculations of magnetic interaction
between a protostar and its accretion disk. By assuming Newtonian
gravity and neglecting the rotation of the star, they showed that the
magnetic interaction can explain the X-ray flares and outflows
observed in protostars.  Later, Goodson, B\"{o}hm, and Winglee;
Goodson \& Winglee (1999); hereafter referred to as GBW99 and GW99,
respectively, showed that the recurring magnetic reconnection creates
periodic outflow along the rotation axis of the disk.  However, these
works have not clearly shown the formation of a magnetic tower.

In this paper, we numerically studied the formation process of
magnetic towers around a weakly magnetized neutron star.  In \S 2 we
present basic equations and models.  Numerical results are shown in \S
3.  \S 4 is devoted for discussion and conclusion.

\section{SIMULATION MODEL}
We solved resistive magnetohydrodynamic equations in cylindrical
coordinates.  For normalization, we use the Schwarzschild radius
$r_{s}=2GM/c^{2}=1$ and the speed of light $c=1$.   General
relativistic effects in the innermost region of the accretion disk
around the neutron star are taken into account by using the
pseudo-Newtonian potential (Paczy\'{n}sky \& Wiita 1980)
$\psi=-1/[2(R-1)]$ where $R=\sqrt{r^{2}+z^{2}}$.  The basic equations
in conservative form are as follows: 
\begin{equation}
{\partial\rho\over \partial t} +
\mbox{\boldmath$\nabla$}\cdot(\rho\mbox{\boldmath$v$}) = 0
\label{continuity}
\end{equation}
\begin{eqnarray}
{\partial\over\partial t}\left(\rho\mbox{\boldmath$v$}\right)
+\mbox{\boldmath$\nabla$}\cdot
  \left(\rho\mbox{\boldmath$v$}\mbox{\boldmath$v$} -
       {\mbox{\boldmath$B$}\mbox{\boldmath$B$}\over 4\pi}\right)
=-\mbox{\boldmath$\nabla$}\left(p+{B^{2}\over 8\pi}\right) -
\rho\mbox{\boldmath$\nabla$}\psi
\label{momentum}
\end{eqnarray}
\begin{eqnarray}
{\partial\over\partial t}\left(\epsilon +{B^{2}\over 8\pi}\right)
+\mbox{\boldmath$\nabla$}\cdot\left[\left(\epsilon +
  p\right)\mbox{\boldmath$v$} + {\mbox{\boldmath$E$}\times
    \mbox{\boldmath$B$}\over 4\pi}\right]
=-\rho\mbox{\boldmath$v$}\cdot\mbox{\boldmath$\nabla$}\psi
\label{energy}
\end{eqnarray}
\begin{equation}
{\partial \mbox{\boldmath$B$}\over \partial t} =
-\mbox{\boldmath$\nabla$} \times \mbox{\boldmath$E$}
\label{induction}
\end{equation}
where $\epsilon=\rho v^{2}/2+p/(\gamma-1)$ is the total energy of
the gas ($\gamma=4/3$), $\mbox{\boldmath$E$} = -\mbox{\boldmath$v$}
\times \mbox{\boldmath$B$} + \eta\mbox{\boldmath$J$}$ and
$\mbox{\boldmath$J$} = (\mbox{\boldmath$\nabla$} \times
\mbox{\boldmath$B$})/4\pi$. The other symbols have their usual
meanings. 

  The resistivity $\eta$ is assumed to be uniform in the entire
  computational domain and parametrized by the magnetic Reynolds
  number $R_{m}=c r_{s}/\eta = 1000$, where we take the characteristic
  speed and length as $c$ and $r_{s}$, respectively.  The resistivity
  is taken to be small enough to ensure the coupling between plasma
  and magnetic fields.  The diffusion time scale of magnetic fields
  $\tau_{d}\equiv {\cal A}/\eta = ({\cal A}/c r_{s})R_{m}$ (${\cal A}$
  is the area of the diffusion region) is much longer than the
  dynamical time scale except in the localized current sheet.  In this
  paper, we assumed a uniform resistivity rather than an anomalous
  resistivity, which incorporates plasma instabilities due to kinetic
  effects, adopted in Hayashi {\it et al.} (1996).  As we show later,
  essential results of Hayashi {\it et al.}'s simulation such as a
  topological change of magnetic field lines and ejection of plasmoids
  are reproduced by numerical simulations with small uniform
  resistivity.

The basic equations are solved by the 2-D axisymmetric MHD code
(Matsumoto {\it et al.} 1996; Hayashi {\it et al.} 1996) based on a
modified Lax-Wendroff scheme with artificial viscosity.

The initial rotating torus is obtained by assuming a polytropic
equation of state $p=K\rho^{1+1/n}$ where $n=3$ and the angular
momentum distribution is $l=l_{0}(r/r_{0})^{a}$ where
$l_{0}=r_{0}^{3/2}/[(r_{0}-1)\sqrt{2}]$ and $a=0.35$.  The subscript
$0$ signifies the values at the density maximum of the torus
$r_{0}=13$ and the density is normalized at this point.  The initial
density distribution of the torus is given by:
\begin{equation}
\rho_{t}=\rho_{0}\left(1-{\gamma\over {v^{2}_{\rm s,0}}}
  {\tilde{\psi}-\tilde{\psi}_{0}\over n+1}\right)^{n}
\end{equation}
%
where $\rho_{0}$, $\tilde{\psi}=\psi+(l/r)^{2}/[2(1-a)]$, and
$v_{\rm s,0}$ are the maximum density of the torus, the effective
potential, and the sound speed of the torus, respectively.  The
thermal energy of the torus is parametrized by $E_{\rm
  th,0}=v^{2}_{\rm s,0}/\gamma|\psi_{0}|$.

Outside the torus, we assume a non-rotating, spherical, and isothermal
hot corona in hydrostatic equilibrium.  The density
distribution of the corona is:
\begin{equation}
\rho_{c}=\rho_{*}\exp{\left[-{\gamma\over
      v^{2}_{\rm s,*}}\left(\psi-\psi_{*}\right)\right]}
\end{equation}
%
The subscript $*$ signifies the values at the surface of the neutron
star $R_{*}=2.8$.  The thermal energy of the corona is parametrized by
$E_{\rm th,*}=v^{2}_{\rm s,*}/\gamma|\psi_{*}|$.

The initial magnetic field of the neutron star is the dipole magnetic
field described by the toroidal component of the vector potential
$A_{\phi}=-m_{*} r/R^{3}$.  The strength of the magnetic field is
defined by the magnetic dipole moment:
\begin{equation}
m_{*}=r_{*}^{3}\sqrt{{2\pi\rho_{*}\over
    \beta_{*}}{v^{2}_{s,*}\over\gamma}}
\end{equation}
parametrized by the value of the plasma $\beta$ defined as the
ratio of gas pressure of the corona to magnetic pressure.

Initial conditions are determined by choosing the non-dimensional
parameters $\rho_{*}/\rho_{0}=0.006$, $E_{\rm th,0}=0.006$, $E_{\rm
  th,*}=0.21$, and $\beta_{*}=2$.  We use $600\times 600$ non-uniform
meshes.  The grid size is uniform ($\Delta r=\Delta z=0.025$) in the
region $0 < r < 10$ and $0 < z < 10$.  Otherwise, it increases with
$r$ and $z$ where $r>10$ or $z>10$.  The size of the entire
computational box is $0\leq r\leq 100$ and $0\leq z\leq 100$.  At the
inner boundary $R=R_{*}$, we imposed an absorbing boundary condition.
The deviation from the initial values are artificially decreased in
the damping layer between $R=R_{*}$ and $R=R_{*}+\Delta R$, where
$\Delta R=0.2$.  We imposed a symmetric boundary condition at the
equatorial plane and a reflecting boundary condition at the rotation
axis.  The outer boundaries are free boundaries where waves can
transmit.

\section{SIMULATION RESULTS}
\begin{figure*}[t]
\centerline{\epsfxsize=\hsize \epsfbox{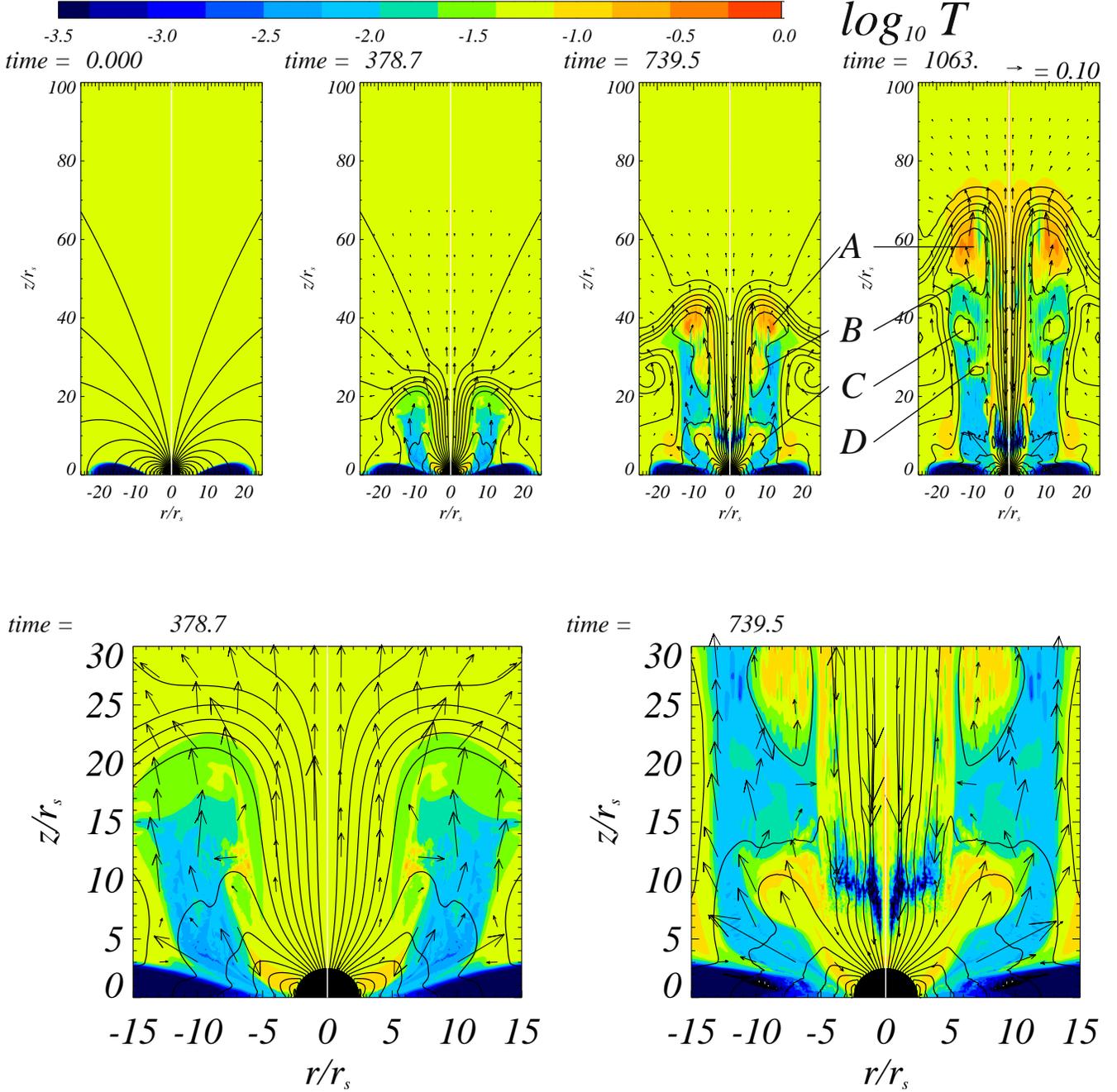}}
\caption{Formation of a magnetic tower, working surface 'A', and
  intermittent ejection of hot bubbles 'B', 'C', and 'D' and their
  time evolution inside the magnetic tower (upper panels).  The color
  contour shows the temperature distribution and arrows show the
  velocity vectors.  The disk material is launched from the inner part
  of the accretion disk (bottom-left panel) to a direction of 60
  degrees from the rotation axis and collimated into the rotation
  axis.  The initial outflow velocity is about 20\% of the speed of
  light.  Hot bubbles are created inside the magnetic loops anchored
  to the disk and the star (bottom-right panel.)}
\label{fig1:eps}
\end{figure*}

Figure \ref{fig1:eps} shows the time evolution of poloidal magnetic
field lines (solid lines) and the distribution of $T=\gamma P/\rho$
(color contours).  The arrows show the velocity vectors. In the upper
panels, the time interval between each panel is about one rotation
period $\tau_{rot} \sim 350r_{s}/c$ at $r=r_{0}$. The bottom panels
enlarge the innermost region.

As the magnetic field lines connecting the neutron star and the disk
are twisted due to the rotation of the disk, they begin to inflate in
the direction 30-60 degrees from the rotation axis. The magnetic
fields cease to splay out when the magnetic pressure balances with the
ambient gas pressure.  Afterwards, the expanding magnetic loops form a
cylindrical tower of helical magnetic fields whose height increases
with time.  In previous MHD simulations of disk-star magnetic
interaction (e.g., HSM96; GBW99 and GW99; Ustyugova et al. 2000), the
tower structure was not so prominent because the ambient gas pressure
was too low to confine the magnetic tower inside the computational
box.

Numerical results indicate that the top interface between the tower
and the external medium propagates with a speed $\sim 0.1c$. This
propagation speed is consistent with that of the theory of magnetic
towers (LB96).  Disk materials are accelerated and launched along the
magnetic tower.  Inside the magnetic tower, current sheets are
formed because magnetic field lines extend upward from the disk and
then go downward to the star.  Magnetic reconnection taking place
in the current sheet injects hot plasmoids intermittently into the
magnetic tower. In figure \ref{fig1:eps}a, symbols 'B', 'C', and 'D'
denote such plasmoids created by magnetic reconnection.  Region 'A' is
the hot region between the top of the magnetic tower and the jet
terminal shock.  Beside this working surface, the jet flow changes its
direction away from the rotation axis and creates backflows. The
radial interface between the magnetic tower and the ambient matter is
stable for the growth of the Kelvin-Helmholtz instability because
magnetic fields stabilize the instability.

\begin{figure*}[t]
\centerline{\epsfxsize=\hsize \epsfbox{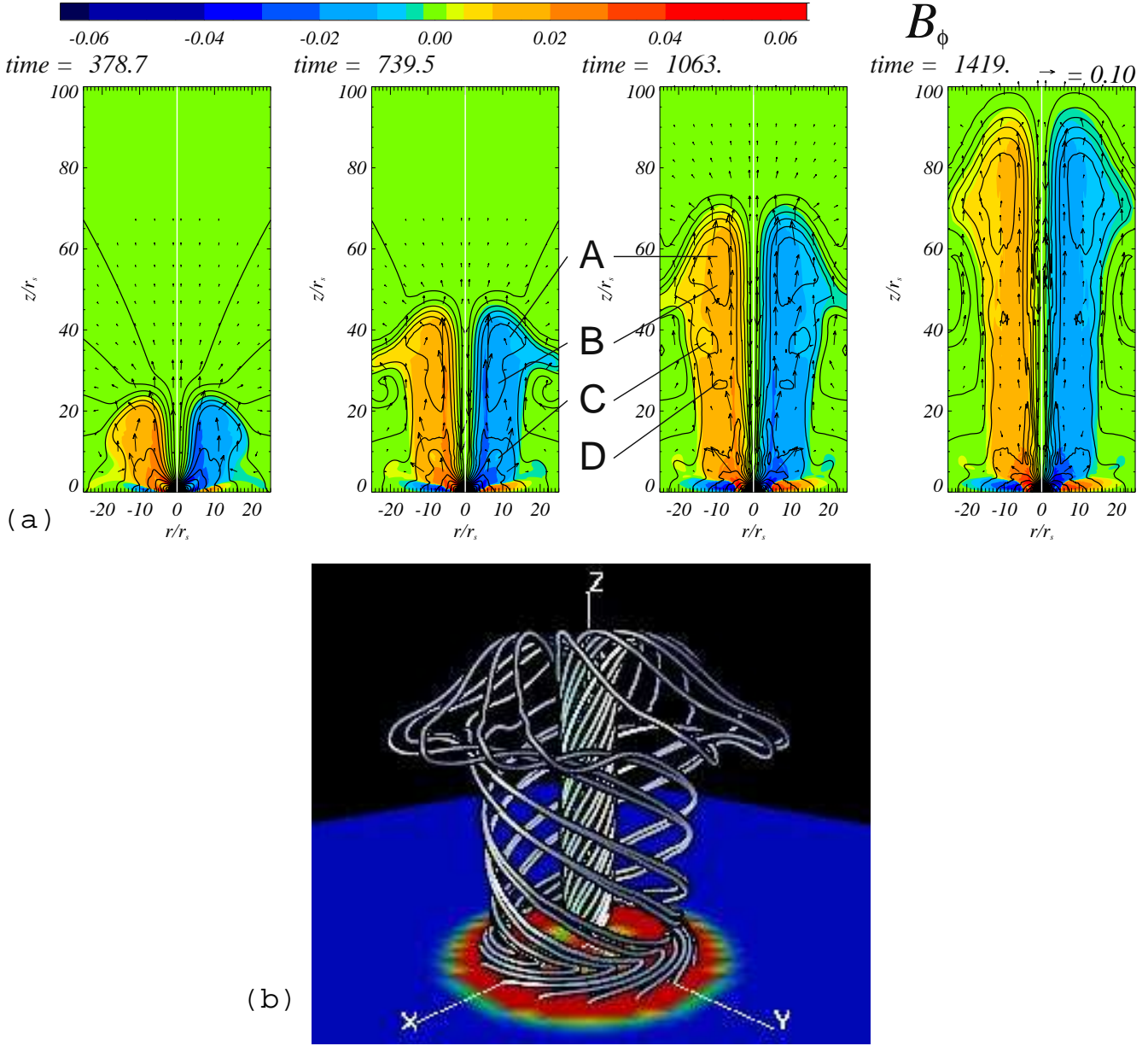}}
\caption{(a) Magnetic field configuration of the magnetic tower in
  {\it r-z} plane (top panels).  Solid curves show  poloidal magnetic
  field lines projected onto this plane.  The color contour shows the
  strength of toroidal magnetic fields and arrows show velocity
  vectors.  The alphabetical symbols indicate the individual magnetic
  islands.  (b) The bottom panel shows the 3-D image of the magnetic
  field lines (solid lines) and density distribution in the equatorial
  plane (color contour).}
\label{fig2:eps}
\end{figure*}

Figure \ref{fig2:eps}a shows the time evolution of the distribution of
toroidal magnetic fields. Figure 2b visualizes the three-dimensional
structure of magnetic field lines at {\it t}=739.5. The magnetic field
lines are strongly twisted due to the rotation of the disk.  Note the
similarity between this figure and the analytical model of magnetic
towers (figure 2 of LB96.)

\begin{figure*}[t]
\centerline{\epsfxsize=\hsize \epsfbox{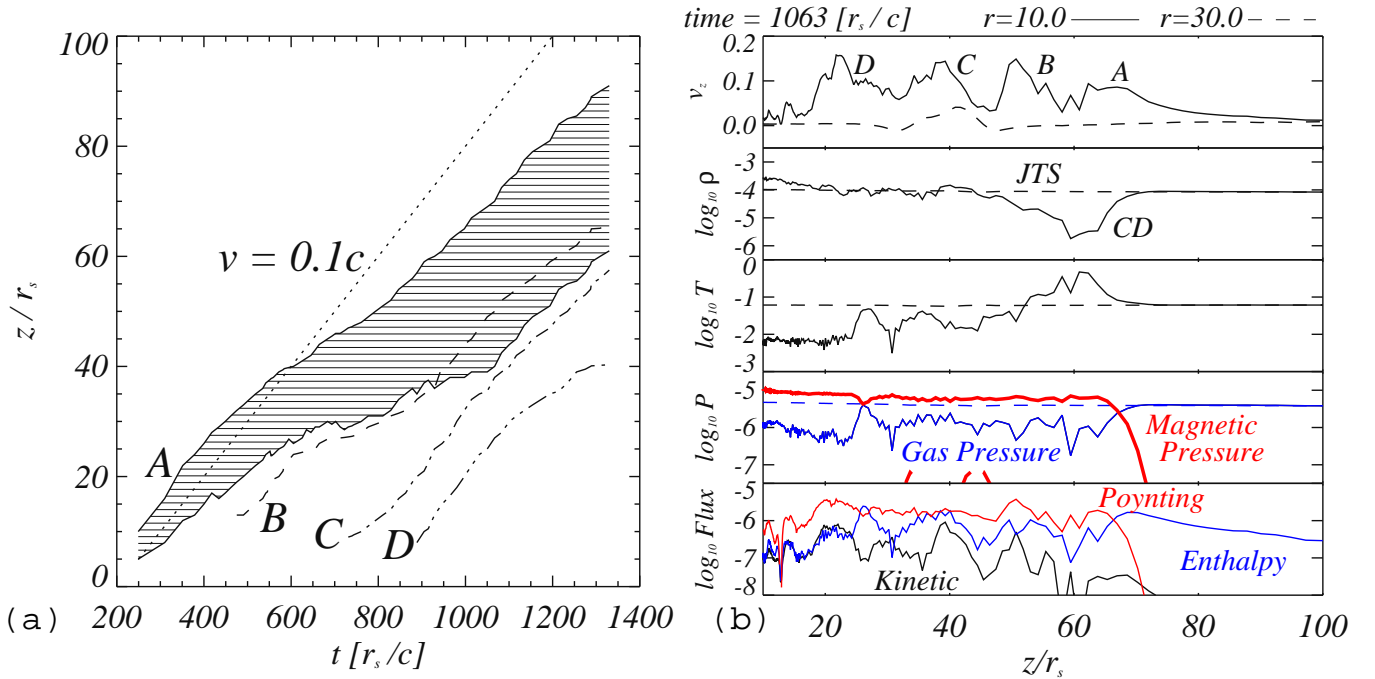}}
\caption{(a) The trajectories of the working surface 'A' and the
  bubbles 'B', 'C', and 'D' in $(t,z)$ plane.  The shaded
  region indicates the working surface 'A'.  At $t=900$, the plasmoid
  'B' collides with 'A' and merges.  The speed of the bubbles is about
  10\% of the speed of light.  The dotted line indicates 10\% of
  the speed of light.  (b) The snapshot of velocity, density,
  temperature, gas pressure (thin), magnetic pressure (thick), and
  distribution of energy flux inside the magnetic tower are shown from
  top to bottom.  Solid and dashed curves indicate the value at $r=10$
  and $r=30$, respectively.  A contact discontinuity (CD) locates
  at $z=67$ and a jet terminal shock (JTS) is placed at $z=52$.}
\label{fig3:eps}
\end{figure*}

Figure \ref{fig3:eps}a shows the trajectories of the working surface of
the magnetic tower 'A' and the bubbles 'B', 'C', and 'D'. The plasmoids
are ejected intermittently with interval of the rotation period around
$5 - 10 r_{s}$. The maximum speed of the plasmoid moving inside the
magnetic tower is about $0.2c$. The plasmoid 'B' catches up with the
working surface 'A' at $t\sim 800$ and merges after $t\sim 900$.

  Figure \ref{fig3:eps}b shows the distribution of velocity,
  density, temperature, gas pressure, magnetic pressure, and the
  vertical energy flux [kinetic energy flux $(\rho v^{2}/2)v_{z}$,
  enthalpy flux $[\gamma p/(\gamma - 1)]v_{z}$, the Poynting flux
  $(\mbox{\boldmath$E$}\times \mbox{\boldmath$B$})_{z}/4\pi$] at
  $r=10$ (inside the magnetic tower) and $r=30$ (outside the magnetic
  tower).  The radius of the tower is about 20. The total pressure of
  the tower slightly exceeds the ambient gas pressure.  It is
  clear that the Poynting flux dominates other fluxes inside the
  jets.

The density distribution has a minimum in the working
surface 'A'. It corresponds to the region between the contact
discontinuity (CD) and the jet terminal shock (JTS). 
  We should remark that the bow shock ahead of the working surface
  is not visible because the sound speed of the ambient plasma
  ($v_{\rm s,*} \sim 0.4$) is larger than the propagation speed of the
  jet.

\section{DISCUSSION}
In this letter, we have demonstrated that highly collimated magnetic
towers are formed along the rotation axis of the accretion disk
surrounding a weakly magnetized neutron star.  We also found that hot
plasmoids created by the intermittent magnetic reconnections are
injected into the magnetic tower (Figure \ref{fig1:eps}).  The
magnetic towers are confined to the radius where the magnetic pressure
of the expanding magnetic loops is comparable to the gas pressure of
the ambient plasma.

Recently, {\it Poynting jet}, in which the energy and angular momentum
are carried predominantly by the electromagnetic field, has been
studied numerically by Romanova {\it et al.} (1998) and Ustyugova {\it
  et al.} (2000).  The magnetic tower jet is a Poynting jet,
  because the Poynting flux dominates the energy flux of the bulk
  flow inside the jet in our simulation (Figure \ref{fig3:eps}b).  The
  electromagnetic extraction of angular momentum from the disk drives
  the accretion of the disk material. It is interesting to compare the
  amount of angular momentum extracted electromagnetically by the jet
  and the angular momentum transported inside the disk due to the
  magnetic turbulence driven by the magneto-rotational instability
  (MRI).  We have to carry out three-dimensional MHD simulations to
  determine the MRI driven angular momentum transport rate
  self-consistently. We intend to report the results of such
  simulations in subsequent papers.  Our simulation demonstrated the
formation mechanism of a semi-relativistic Poynting jet around a
neutron star.

  We can identify the jet terminal shock (JTS) behind the working
  surface of the jet.  Our numerical results show that a hot, low
  density region is created between the JTS and the discontinuity (CD)
  separating the jet material and the ambient plasma.  The structure
  of the working surface of the jet is qualitatively consistent with
  those in previous MHD simulations of jet propagation assuming
  force-free magnetic fields initially uniform in the $z$-direction
  (e.g., Todo et al. 1992), although the detailed structure is
  different because our numerical simulations started with a dipole
  magnetic field.

We found that hot plasmoids are injected intermittently into the
magnetic tower.  The injection speed of the blobs is the order of the
local Alfv\'{e}n speed of the reconnection region ($v_{A}\sim 0.2c$),
which is comparable to the rotation speed of the innermost region of
the disk.  Since this speed is faster than the propagation speed of
the working surface of the jet, the blobs catch up with the working
surface and release magnetic energy.  Magnetic reconnection taking
place in the core and in the working surface could generate high
energy electrons which emit synchrotron radiation.

  Let us discuss the correlation between the radio flare of the
  core and the lobe observed in Sco X-1.  Magnetic reconnection
  taking place in the core will eject hot plasmoids, which
  subsequently release magnetic energy again at the working surface of
  the magnetic tower.  This event can be observed as the lobe flares
  correlated with the core flares.  In our simulations, however, the
  maximum speed of the hot plasmoids ($\sim 0.2c$) is smaller than the
  speed estimated from the time lag between the core flare and the
  lobe flare.  This discrepancy may be resolved if the Alfv\'{e}n
  speed inside the magnetic tower is faster than that attained in our
  simulation.  This happens when the plasma density inside the
  magnetic tower is lower than the ambient plasma.  Plasma heating due
  to magnetic reconnection helps increasing the Alfv\'{e}n speed and
  ejection speed of hot bubbles.

  We should point out the limitation of using non-relativistic
  MHD.  Although we take into account general relativistic effects by
  using a pseudo-Newtonian potential, we solved Newtonian MHD
  equations by neglecting special relativistic corrections.  By
  extending our work to fully general relativistic MHD, we will be
  able to simulate models having Alfv\'{e}n speed closer to the light
  speed.

The work presented here is the first numerical simulation which
clearly shows the formation process of the axisymmetric magnetic
tower and semi-relativistic jets.  We expect that
  semi-relativistic jets observed in some XRBs are driven by the
  magnetic interaction between the dipole magnetic field of a weakly
  magnetized neutron star and its accretion disk.  3-D simulations
are necessary to study the stability of magnetic towers against
non-axisymmetric perturbations.  We would like to report the results
of such simulations in the near future.

\acknowledgments

The authors would like to thank K. Shibata, S. Mineshige, and D. Meier
for valuable discussions.  One of us (YK) appreciate stimulating
discussion with R. E. Pudritz in Les Houches Euro Summer School 2002.
Numerical computations were carried out on VPP5000 at the Astronomical
Data Analysis Center of the National Astronomical Observatory, Japan
(myk18b).  The work is supported in part by ACT-JST of Japan Science
and Technology corporation and Grants-in-Aid of the Ministry of
Education, Culture, Sports, Science, and Technology of Japan
(15037202, RM).

\clearpage


\end{document}